\documentclass{PHYEAUTH}
\usepackage{graphicx}
\usepackage{amsmath}
\usepackage{amssymb}

\begin{document}

\begin{frontmatter}

\title{In-plane Magnetic Field Dependent Magnetoresistance\\ 
of Gated  Asymmetric Double Quantum Wells}

\author[address1]{Yu. Krupko},
\author[address1]{L. Smr\v cka\thanksref{thank1}},
\author[address1]{P. Va\v sek},
\author[address1]{P. Svoboda},
\author[address1]{M. Cukr},
and
\author[address2]{L.Jansen}

\address[address1]{Institute of Physics ASCR, Cukrovarnick\'a 10, 162 53
Praha 6, Czech Republic}
\address[address2]{Grenoble High Magnetic Field Laboratory,
Bo\^{\i}te Postale 166 , 38042 Grenoble Cedex 09, France}

\thanks[thank1]{Corresponding author.
E-mail: smrcka@fzu.cz}

\begin{abstract}
We have investigated experimentally the magnetoresistance of strongly
asymmetric double-wells. The structures were prepared by inserting a
thin Al$_{0.3}$Ga$_{0.7}$As barrier into the GaAs buffer layer of a
standard modulation-doped GaAs/Al$_{0.3}$Ga$_{0.7}$As
heterostructure. The resulting double-well system consists of a nearly
rectangular well and of a triangular well coupled by tunneling through
the thin barrier.  With a proper choice of the barrier parameters one
can control the occupancy of the two wells and of the two lowest
(bonding and antibonding) subbands. The electron properties can be
further influenced by applying front- or back-gate voltage.  
\end{abstract}

\end{frontmatter}
\section{Introduction}
A magnetic field $B_{\|}$ applied parallel to the
quasi-two-dimensional systems of electrons confined in double-well
structures is known to couple strongly to the electron orbital motion
and to change dramatically the electron energy spectra. The
magnetoresistance oscillation observed on coupled double quantum wells
represents a striking manifestation of van Hove singularities in the
$B_{\|}$-dependent density of states, corresponding to the
depopulation of the antibonding subband at a critical field $B_{\|} =
B_{c,1}$, and to the splitting of the Fermi sea into two separated
electron sheets at a second critical field $B_{c,2}$
\cite{si,ku,ju,mak}. There can also be the third critical field
$B_{c,3}$ \cite{sv} at which the system returns to the single layer
state due to depletion of the triangular well.
\section{Experiments}
Strongly asymmetric double-well structures were prepared (by MBE
method) by inserting a thin Al$_{0.3}$Ga$_{0.7}$As barrier into the
GaAs buffer layer of standard modulation-doped
GaAs/Al$_{0.3}$Ga$_{0.7}$As heterostructures. The resulting
double-well system consists of a nearly rectangular well and of a
triangular well.  The bonding and antibonding subbands are formed by
tunnel-coupling of the ground states in individual wells.

The resulting separation of the bonding and antibonding levels depends
on the position and on the thickness of the barrier.  The minimum
separation is achieved if the barrier is located close to the center
of mass of the two-dimensional electron layer.  With the barrier
closer to or farer from the interface the bonding and antibonding
subbands become more separated.

Two types of structures with similar bonding and antibonding energy
levels can be prepared: (i) The barrier is far from the interface and
only a small part of electrons is in the triangular well. (ii) The
barrier is close to the interface and the majority of electrons is in
the triangular well.
 
The occupancy of the wells can be further modified by applying front-
or back-gate voltage. The back-gate voltage will influence strongly
the structures of the first type while the samples of the second type
will be more sensitive to the front-gate voltage.

The response of a structure to the applied in-plane magnetic field
depends on its type.

For both types the Fermi contours are two concentric circles at
$B_{\|} = 0$.  The in-plane field $B_{\|}$ induces a transfer of
antibonding electrons to the bonding subband and the shift of the
centre of mass of the electron layer closer to the interface
\cite{mak}.

The Fermi contour of the antibonding subband changes its shape,
shrinks and, at a critical field $B_{c,1}$, disappears. All electrons
occupy only the bonding subband and form a wide single layer. The
magnetoresistance approaches a pronounced minimum as the density of
states (DOS) is halved and, therefore, also the electron scattering
rate.

Upon further increasing $B_{\|}$, the system undergoes a transition
from a single-layer to a bilayer state.  A neck in the peanut-like
Fermi contour becomes narrower and, at $B_{\|} = B_{c,2}$, the contour splits
into the  Fermi lines of two independent electron sheets
localized in the rectangular and triangular wells.  At this second
critical field, the DOS has a logarithmic singularity and the
magnetoresistance reaches a sharp maximum.

\begin{figure}[ht]
\begin{center}
\includegraphics[angle=-90,width=\linewidth]{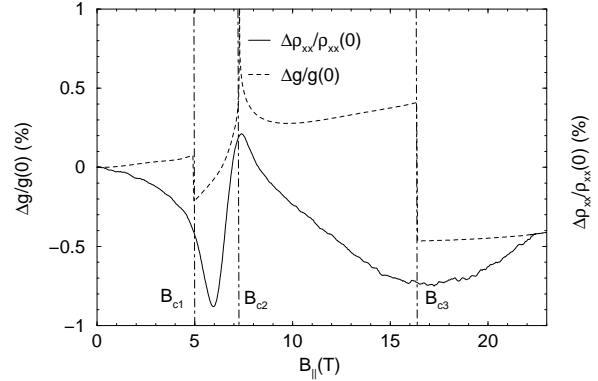}
\end{center}
\caption{Magnetoresistance $\Delta \rho/\rho(0)$ together with the
calculated density of states $\Delta g/g(0)$.  Sample A.}
\label{Fig1}
\end{figure}

Above $B_{c,2}$, the shift of the electron sheet towards the interface
continues. In the first type of structures all electrons are
transfered into the rectangular well at a third critical field
$B_{c,3}$ and the 2D system re-enters the single-layer state
\cite{mak}. An example of the magnetoresistance trace  and the calculated DOS
is in Fig.~\ref{Fig1}.

The aim of this paper is to study the influence of front- and
back-gate voltages on the above mentioned critical fields.

Two samples grown by MBE were studied. The Sample A with a wide
rectangular well and the Sample B with a narrow rectangular well. In
Table~\ref{tab}, the width of the rectangular well is denoted by $d$,
the width of the barrier is $w$. The experiments were carried out at
temperature $0.4$~K, using low-frequency (13 Hz) $ac$ technique.

\begin{table}[hb]
\caption{}
\begin{tabular*}{\linewidth}{ c c c c c c c c }
\hline
\hline
\,\,Sample & $d$ & $w$ &\,\,\,\,&  $N_b$ & $N_a$ & $N$ & $\mu$\\
       & (nm) & (nm)&& &$\rm(10^{11}cm^{-2})$& &$\,\,\rm(10^5cm^2/Vs)$\\
\hline
  A    & 10   & 3.4 && 3.02 & 0.25 & 3.27& 1.16\\
  B    &  7   & 1.7 && 2.64 & 0.57 & 3.21 & 0.97\\
\hline
\hline
\end{tabular*}

\label{tab}
\end{table}
The presented characteristics of 2DEG have been extracted from the
data taken in perpendicular magnetic fields. The total electron
concentration $N$ and the electron mobility $\mu_H$ were calculated
from the low-field Hall and zero-field resistivities, respectively.
The partial concentrations of electrons in the subbands, $N_b$ and
$N_a$, were determined from the period of Shubnikov-de Haas (SdH)
oscillations.

\section{Results and discussion} 
\subsection{Sample A}
The Sample A has a 10 nm wide rectangular well and only the minority
of electrons is behind the barrier in the triangular well.  Therefore,
it is a good candidate for application of the back-gate voltage. (The
magnetoresistance  obtained for zero gate voltage is shown  in
Fig.~\ref{Fig1}.)
%
\begin{figure}[t]
\begin{center}
\includegraphics[angle=-90,width=0.9\linewidth]{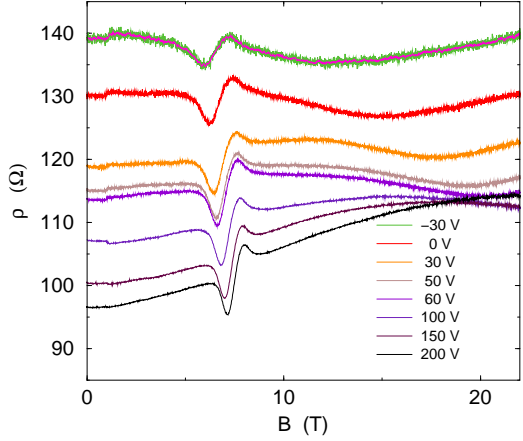}
\end{center}
\caption{Magnetoresistance traces for Sample A  measured in
$B_{\|}$ for different back-gate voltages.}
\label{Fig2}
\end{figure}

Experimental curves for a range of back-gate voltages are presented in
Fig.~\ref{Fig2}.  The back gate has strong effect on the electrons of
low-populated triangular well. While the critical fields $B_{c,1}$ and
$B_{c,2}$ do not depend substantially on the applied gate voltage, the
third critical field $B_{c,3}$ (at which the system returns to the
single layer state) is very sensitive to the change of the total
concentration $N$ induced by the back-gate voltage. An increase of $N$
due to positive back-gate voltages leads to a shift of $B_{c,3}$ to
higher magnetic fields. At higher possitive back-gate voltages there are too
many electrons in the triangular well and the minimum in
magnetoresistance corresponding to $B_{c,3}$ is beyond the available
magnetic fields.

\subsection{Sample B}
The Sample B has a 7 nm wide rectangular well.  The application of
the front-gate voltage has even more dramatic effect on its electronic
structure than the back-gate on the Sample A. 
%
\begin{figure}[t]
\begin{center}
\includegraphics[angle=0,width=\linewidth]{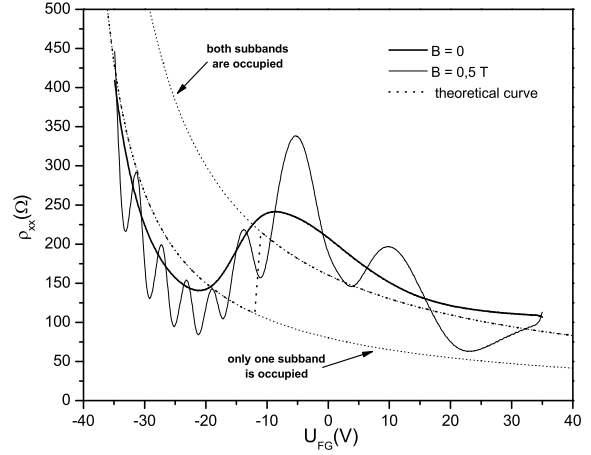}
\end{center}
\caption{The resistivity of Sample B dependent upon the
front-gate voltage.}
\label{Fig3}
\end{figure}

\begin{figure}[hb]
\begin{center}
\includegraphics[angle=0,width=\linewidth,height=0.8\linewidth]{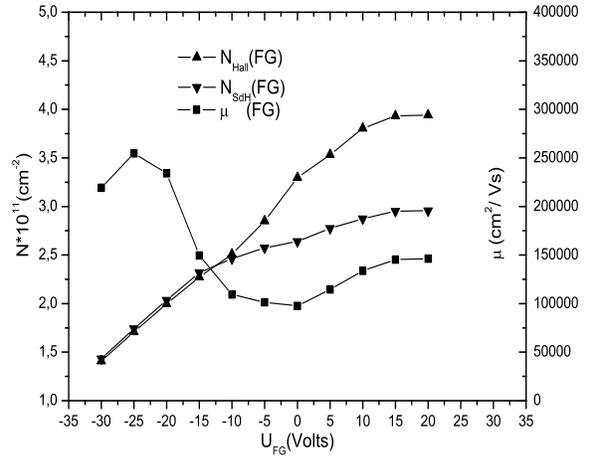}
\end{center}
\caption{The dependence of partial concentrations and of 
the mobility
on the gate voltage. Sample B.}
\label{Fig4}
\end{figure}
%
As it can be deduced from resistivity curves in Fig.~\ref{Fig3},
the negative gate voltage reduces the electron concentration $N$ and
is able to empty the antibonding subband without application of the
in-plane magnetic field. At the same time, the sheet of electrons is
repulsed from the interface behind the barrier, i.e into the triangular
well.

Qualitatively, the variation of the resistivity is expected to be
proportional to the scattering rate (which is approximately doubled at
the crossover from the single-subband to the double-subband occupancy)
and inversely proportional to the concentration $N$. This is
schematically sketched in Fig.~\ref{Fig3} by dotted lines.  The
zero-field resistance exhibits an oscillation corresponding to this
scheme with the crossover around $U_{FG} = -13$~V. The single-subband
to the double-subband occupancy is confirmed by a sudden change of
period of SdH oscillations observed on the magnetoresistanence trace
measured at $B_{\perp}=0.5$~T. The fast oscillations for $U_{FG} < -13$~V
are attributed to the single-subband occupancy, the slow oscillation
above $U_{FG} = -13$~V are due to $N_b$ electrons in the bonding subband.

The quantitative results for the dependence of partial concentrations
and of the mobility on $U_{FG}$, obtained from low field
magnetoresistance measurements, are summarized in
Fig.~\ref{Fig4}. Here we assume that the Hall voltage determines the
total concentration, $N = N_{Hall}$, and only electrons in the
bonding subband are responsible for the SdH oscillations, $N_b =
N_{SdH}$, when two subbands are occupied.
\begin{figure}[h]
\begin{center}
\includegraphics[angle=0,width=\linewidth,height=\linewidth]{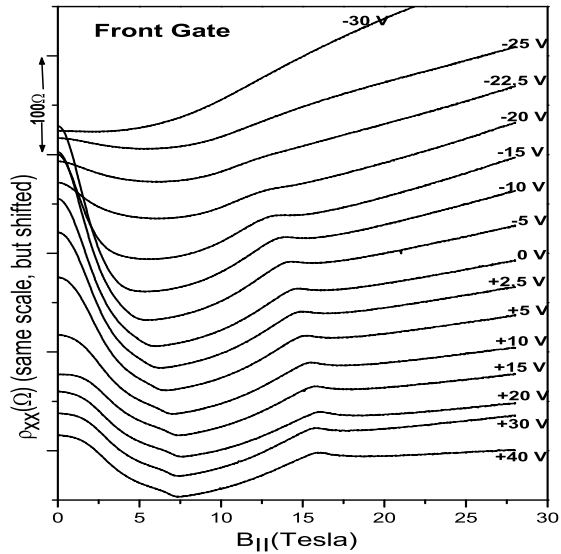}
\end{center}
\caption{Magnetoresistance traces for Sample B measured in
parallel field configuration for different front-gate voltages. The
gate-voltage values are printed near the corresponding curves.}
\label{Fig5}
\end{figure}

Magnetoresistance traces for Sample B, measured in parallel field
configuration for different front-gate voltages, are displayed in
Fig.~\ref{Fig5}. Positive gate voltages shift the critical fields
$B_{c,1}$ and $B_{c,2}$ to higher values as $N$ (and mainly $N_a$)
increase. As the negative $U_{FG}$ reduces $N_a$, $B_{c,1}$ shifts to
lower fields and the corresponding singularity becomes less
pronounced. The same is true also for $B_{c,2}$. While $B_{c,1}$
disappears as $N_a\rightarrow 0$, we attribute the suppression of the
logarithmic singularity at $B_{c,2}$, related to the
single-layer/double-layer transition, to the fact that, at most
negative $U_{FG}$, no neck is formed on the Fermi contour, the Fermi
contour does not split into two parts and no independent electron
sheet is formed in the rectangular well.

\section{Conclusions}
The magnetic field oriented in parallel to the 2D electron system
induces a deformation of the Fermi contours corresponding to bonding
and antibonding subbands, depopulation of the higher occupied subband
at the critical field $B_{c,1}$, and the transition into the decoupled
bilayer at the critical field $B_{c,2}$.

With the small concentration of electrons in the triangular well, we
can observe the third critical field $B_{c,3}$ at which the triangular
well is emptied and the system returns to the single layer state. The
critical field $B_{c,3}$ is strongly dependent on the back-gate
voltage.

The  negative front-gate voltage can  empty  the antibonding
subband and suppress $B_{c,1}$ and $B_{c,2}$ in samples with small
concentration of electrons in the rectangular well. In that case, no
single-layer/double-layer transition occurs.

\section*{Acknowledgements}
This work has been supported by the Grant Agency of the Czech
Republic under Grant No. 202/01/0754, by the French-Czech project
Barrande 99011, and by the European Community project ``Access to
Research Infrastructure action of the Improving Human Potential
Programme''.

\enlargethispage*{0.5cm}

\end{document}